\documentclass[useAMS,usenatbib]{mn2e}
 \usepackage[dvips]{graphicx}
 \usepackage[english]{babel}

\title[Measuring Galaxy morphology in clusters]
{Quantifying galactic morphological transformations in the cluster environment}
\author[Cervantes-Sodi, Park, Hernandez \& Hwang]{B. Cervantes-Sodi$^{1}$\thanks{E-mail: bcsodi@kasi.re.kr, cbp@kias.re.kr, xavier@astroscu.unam.mx and hoseong.hwang@cea.fr},
Changbom Park$^{2}$,
X. Hernandez$^{3}$ and Ho Seong Hwang$^{4}$\\
$^{1}$Korea Astronomy and Space Science Institute, 61-1 Hwaam-dong, Yuseong-gu, Daejeon 305-348, Korea \\
$^{2}$Korea Institute for Advanced Study, Dongdaemun-gu, Seoul 130-722, Korea\\
$^{3}$Instituto de Astronom\'\i a,
Universidad Nacional Aut\'onoma de M\'exico
A. P. 70--264,  M\'exico 04510 D.F., M\'exico \\
$^{4}$CEA Saclay/Service d'Astrophysique, F-91191 Gif-sur-Yvette, France\\
}
\begin{document}

\date{In original form 2010 August 16}

\pagerange{\pageref{firstpage}--\pageref{lastpage}} \pubyear{2009}

\maketitle

\label{firstpage}

\begin{abstract}

We study the effects of the cluster environment on galactic morphology by  
defining a dimensionless angular momentum parameter $\lambda_{d}$,
to obtain a quantitative and objective measure of galaxy type. The use of this
physical parameter allows us to take the study of morphological transformations in
clusters beyond the measurements of merely qualitative parameters, e.g. S/E ratios,
to a more physical footing. 
To this end, we employ an extensive Sloan Digital Sky Survey sample (Data Release 7),
with galaxies associated
with Abell galaxy clusters. The sample contains 121 relaxed Abell clusters and over
51,000 individual galaxies, which guarantees a thorough statistical coverage over a wide
range of physical parameters. We find that the median $\lambda_{d}$ value tends to
decrease as we approach the cluster center, with different dependences according
to the mass of the galaxies and the hosting cluster;
low and intermediate mass galaxies showing a strong
dependence, while massive galaxies seems to show,
at all radii, low $\lambda_{d}$ values.
By analysing trends in $\lambda_{d}$ as functions of the nearest neighbour
environment, clustercentric
radius and velocity dispersion of clusters, we can identify clearly the leading physical
processes at work. We find that in massive clusters ($\sigma>700$ km/s), the interaction
with the cluster central region dominates,
whilst in smaller clusters galaxy-galaxy interactions are chiefly responsible
for driving galactic morphological transformations.

\end{abstract}


\section{Introduction}

Galaxy evolution and the effects of environment upon it, is one of the major areas
of modern research in extragalactic astronomy, being the cluster environment of
great interest given the variety of processes present there.

Most galaxies reside in galaxy clusters, where the highest
volume densities of galaxies are found, offering a unique laboratory to study
the effects of the environment on different galactic properties. Given the
high density of galaxies in these systems, any environmental dependence
should be more pronounced, in contrast with the field. One such dependency
is the morphology-density relation (Dressler 1980; Treu et al. 2003),
in which the average morphology of galaxies changes as a function
of the local environment. Subsequent studies (e.g., Whitmore \& Gilmore 1991; Whitmore,
Gilmore \& Jones 1993) have shown that the morphology-clustercentric
radius relation seems tighter than the morphology-local density relation,
and explained the origin of this relation as a result of galaxy
interactions with the mean tidal field of the clusters, but
the exact nature of the processes responsible for establishing these relations
are still open to debate.

In galaxy clusters, a possible mechanism for the morphology transformation
is ram pressure stripping, proposed by Gunn \& Gott (1972). According to this model,
the ram pressure of the intracluster medium sweeps away gas from the galactic disks
as they fall at high speed into the cluster, making the remaining galaxy take an
early type appearance. However, the entire structure transformation, such as the
thickening of the discs or the large bulge-to-disc ratios, can not be explained by this process alone
(for a discussion see, e. g., Dressler 1984).

Interactions between galaxies are another agent of morphological
transformations acting through different mechanisms, such as 
galaxy harassment via high-speed encounters,
specially effective in transforming faint Sc-Sd galaxies into
dSphs and even LSBs (Moore et al. 1996); galaxy mergers that exhaust
a large amount of the interstellar gas of interacting gas-rich spirals
(Fujita 1998),
or dramatically destroy galactic discs and thereby convert spiral galaxies into
ellipticals and lenticulars (Toomre \& Toomre 1972).

Finally, tidal interactions with the cluster as a whole are also a source of morphological
transformations. As galaxies fall into the cluster, the outer part of their haloes
will be stripped by the tides produced by the cluster potential (White \& Rees 1978;
Mamon 1992),
and, as soon as they enter the cluster, will experience interactions with the
substructure within the cluster. The tidal field interaction with the cluster
can also induce collisions of disc gas clouds that trigger enhanced nuclear
and disc activity (Byrd \& Valtonen 1990), while the compressive effect of the interaction
within the core of the cluster increases the random velocities of the components
of the disc with a corresponding thickening of the disc (Valluri 1993).

The complete explanation probably includes the joint participation of several of these
processes for transforming the overall structure of galaxies in the cluster
environment.

Among galaxy morphology, many physical parameters have 
been reported to depend on the environment (see Blanton \& Moustakas 2009
for a review). Using a sample of galaxies with
visually classified morphologies, Skibba et al. (2009) found that much of the
morphology-density relation is due to the relation between colour and density. 
As for the star formation rate, Kauffmann et al. (2004) found that it presents the
most sensitive dependence on environment between the properties analyzed on their study
and, Deng (2010) showed that this dependence prevails even at fixed morphology, being
the dependence for late-type galaxies stronger than for early-types.
The case of AGN activity seems to follow two different behaviours in low
and high density environments, in low density environments the fraction of AGN-host
galaxies tends to increase as the local density increases, but it decreases
in high density environments (Lee et al. 2010), specially in the inner
regions of clusters where the high velocity dispersion of galaxies
prevents long term interactions and mergers which are the principal
driving mechanisms in triggering AGN activity (Best 2004).
The decrease in luminosity for high density environments
(Park, Gott \& Choi 2008; Robotham, Phillipps \&
de Propris 2010), and the dependence of the petrosian radius and central
surface brightness on the clustercentric distance (Coenda \& Muriel 2009)
have also been studied, as well as relations between different galaxy properties
such as the size-luminosity relation that appears to be independent of the local
environment, sensitive mostly to galactic morphology (Coenda \& Muriel 2009; Nair, van den Bergh \& Abraham 2010), 
although if morphological transformations are taking place
in high density environments, the issue could be more complicated;
the colour magnitude relation being an other empirical relation
that shows a trend for systematically
bluer galaxy colours with increasing projected radius from the centre of the cluster,
accompanied by a decreased on the internal scatter (Terlevich, Caldwell \&
Bower 2001. 

However, there is a debate about the role played
by the large-scale background density, in comparison to the local
density, in determining galaxy morphology. Recently,
Park, Gott \& Choi (2008) and Park \& Choi (2009), found that some galaxy properties,
like morphology and luminosity, strongly depend on the distance and morphology
of the nearest neighbour galaxy, even at fixed large-scale background density,
with important effects at a characteristic scale given by the virial radius
of the nearest neighbour (see also Hwang \& Park 2009).
Outside this scale, galaxy morphology still depends on the
distance to the nearest galaxy, but showing no dependence on the morphology of
the companion. This indicates that galaxies interact hydrodynamically when they
are close enough, though the interaction is purely gravitational at larger separation
distances. Given the statistical correlation of galaxy-galaxy interactions with
the large-scale background density, plus
the additional influence of the gravitational interaction, galaxy-cluster
potential, an underlying relation between structural
parameters of cluster galaxies and clustercentric radius is expected.

Park \& Hwang (2009; henceforth PH09) studied galaxy properties as a function of
both; clustercentric radius and nearest neighbour galaxy distance, and found that
the structural and kinematic parameters of late-type galaxies such as the
concentration index and the central velocity dispersion depend mainly on the
clustercentric radius, but the star formation activity indicators like the u-r colour
and the equivalent width of H$\alpha$ line, depend predominantly on the nearest
neighbour distance. The former finding may indicate important effects,
or repeated interactions with neighbouring galaxies, or
galaxy-cluster potential interactions on the internal structure of intermediate
mass galaxies, in contradiction with the predictions of the ram pressure stripping
scenario.

Intimately linked with the morphology and present day structure of galaxies
is the galactic spin. Theoretical studies of galaxy formation and
evolution, show the major role played by this physical parameter in determining
properties such as the star formation efficiency of disc galaxies 
(Dalcanton, Spergel \& Summers 1997; van den Bosch 1998),
the bulge  to disc ratio (van den Bosch 1998),
color gradients (Prantzos \& Boissier 2000), and gas content (Churches,
Nelson \& Edmunds 2001, Boissier et al 2001), as well as playing a major
role in explaining the dispersion about many empirical relations such as the Tully-Fisher
relation (Koda, Sofue \& Wada 2000) and specially the size-dependent relations
such as the size-luminosity relation
(Courteau et al. 2007), to mention just a few. All these
parameters, related with the overall morphology of the galaxies, show, to a certain
degree, dependences with the environment, specially for cluster systems.

Motivated by theoretical studies (e.g. Fall \& Efstathiou 1980; van der Kruit 1987; 
Flores et al. 1993; Firmani et al. 1996; Mo, Mao \& White 1998; Prantzos \& Boissier 2000),
Hernandez \& Cervantes-Sodi (2006) and later Cervantes-Sodi \& Hernandez
(2009), showed, with different samples of well studied galaxies, how the
spin parameter correlates with several galactic properties, like the
bulge-to-disc ratio, disc thickness, colours, metallicity and gas abundance.
The above, using a simple estimate of the actual galactic $\lambda$ parameter,
based on easily available observational quantities.
The galaxies analysed showed a systematic increase of the spin when moving
from early to late type spiral galaxies, and in Hernandez et al. (2007),
a match is found for the Hubble type with the angular momentum $\lambda$ parameter
when comparing against a colour versus colour gradient criteria
(Park \& Choi 2005), where the segregation by spin coincides with the segregation by Hubble type,
as visually assigned for a large sample of spiral galaxies from the SDSS.
This demonstrated the plausibility of the identification of the $\lambda$
galactic spin parameter as the principal physical parameter driving galactic
morphology, and has recently been used as such by e.g. Berta et al. (2008)
and Gogarten et al. (2010). The above, under the approximate estimate for $\lambda$
proposed in Hernandez \& Cervantes-Sodi (2006), and shown to be an unbiassed
estimator for the true $\lambda$ parameter through extensive comparisons with 
simulated galaxies in Cervantes-Sodi et al. (2008).

The purpose of this paper is to investigate the possible dependence of the
galactic angular momentum on the cluster environment, given the tight link between
$\lambda$ and morphology, and the morphology-clustercentric radius
relation. In going to a physically well defined and objective parameter to measure
'type', we can complement existing qualitative studies working on more subjectively
assigned 'type' parameters and perform for example, meaningful comparisons with models and
physical identification of the main causes of the evolution seen in clusters.

The outline of this paper is as follows. Section 2 presents the derivation
of the dimensionless angular momentum parameter used to characterize the morphology
of the galaxies in our sample, in Section 3
we introduce the sample used in the study, followed by the results and a discussion
in Section 4 and our general conclusions in Section 5.

\section{Estimation of the spin from observable parameters}

Traditionally, the galactic angular momentum is characterized by the $\lambda$ parameter
as introduced by Peebles (1969); 

\begin{equation}
\label{Lamdef}
\lambda = \frac{L \mid E \mid^{1/2}}{G M^{5/2}}
\end{equation}

where $E$, $M$ and $L$ are the total energy, mass and angular momentum of the configuration, 
respectively. In 
Hernandez \& Cervantes-Sodi (2006) we derived a simple estimate of total $\lambda$ for dark halos
hosting disc galaxies 
in terms of observational parameters, based on two simple hypothesis: that the specific angular
momentum of dark matter and baryons are equal, and a constant small baryonic fraction, for systems
where the total energy and angular momentum are dominated by the dark matter component. The problem with
galaxies residing in clusters is that the diversity of the interactions taking
place in that environment can drastically
change the baryonic fraction and disturb the baryonic and dark matter angular momentum profiles
in different ways, invalidating these two previous hypothesis. For example, tidal forces
can shear off most of the original dark halo of a given galaxy. We must therefore use 
an angular momentum parameter which focuses on the dynamics of the baryonic galaxy.
We retain the quantitative and objective nature of the study, 
and account for the angular momentum focusing only on the baryonic component to define a
baryonic dimensionless angular momentum parameter $\lambda_{d}$.

In our model, we consider a disc for the baryonic component of the galaxy with
an exponential surface mass density $\Sigma(r)$;

\begin{equation}
\label{Expprof}
\Sigma(r)=\Sigma_{0} e^{-r/R_{d}},
\end{equation} 

where $r$ is a radial coordinate and $\Sigma_{0}$ and $R_{d}$ are two constants which are allowed 
to vary from galaxy to galaxy, and assume the presence of a dark matter halo responsible for
establishing a rigorously flat rotation curve $V_{d}$ throughout the disc.

From equation~\ref{Expprof}, the total disc mass is

\begin{equation}
\label{Discmass}
M_{d}=2 \pi \Sigma_{0} R_{d}^{2},
\end{equation}

that combined with our flat rotation curve leads to an angular momentum of
$L_{d}=2V_{d}R_{d}M_{d}$.
Assuming the disk to be a virialized dynamical system, the total energy
can be obtained from the total kinetic energy, estimated as arising merely
from the dominant rotation. In this case, the kinetic energy
of the disc is $T_{d}=M_{d}V_{d}^{2}/2$.

These assumptions allow us to express $\lambda_{d}$ as

\begin{equation}
\label{Lambdad}
\lambda_{d}= \frac{L_{d} \mid T_{d} \mid^{1/2}}{G M_{d}^{5/2}} = \frac{2^{1/2} V_{d}^{2} R_{d}}{G M_{d}}.
\end{equation}

Finally, we introduce a baryonic disc Tully-Fisher (TF)
(e.g  Gurovich et al. 2004, McGaugh et al. 2005) relation: 
$M_{d}=A_{TF} V_{d}^{3.5}$, to replace the dependence on mass for a dependence
on $V_{d}$, to obtain:

\begin{equation}
\label{Ld}
\lambda_{d} = \frac{2^{1/2} R_{d}}{G A_{TF} V_{d}^{3/2}}.
\end{equation}

The presence of an intrinsic dispersion in the Tully-Fisher relation dominates the budget of 
our internal errors, and results in a 25\% uncertainty associated to each individual $\lambda_{d}$
estimate.
Basically, equation~\ref{Ld} is the same expression we derived in Hernandez \& Cervantes-Sodi
(2006) to estimate the traditional $\lambda$ spin parameter for dark matter haloes hosting
disc galaxies, divided by the baryonic fraction of the galaxy; this because to obtain the total
spin parameter we were assuming angular momentum conservation for both components, and a
constant baryonic fraction, linking the specific angular momentum and mass of the dark
matter to those quantities of the baryonic component, while in our current work, we do
not attempt to constrain the physical characteristics of the halo, but just
consider its participation in establishing the flat rotation curve throughout the disc.
The parameter $\lambda_{d}$ is hence not a $\lambda$ parameter in the sense of the definition
of equation~\ref{Lamdef}, but merely an estimate of a dimensionless angular momentum for a galactic disk,
expected to correlate tightly with all type defining properties, as already pointed out 
in the introduction.

As can be seen from equation (5), our estimate of $\lambda_{d}$ is a measure of disk scale length 
and rotational velocity or total luminosity by introducing a Tully-Fisher relation (TF), and therefore, 
equivalent to a measure of the size-luminosity relation. However, we prefer to
present our results in terms of a dimensionless disk angular momentum, given the clear interpretation
and well established theoretical expectations from galactic structure models in terms of this parameter
(Dalcanton, Spergel \& Summers 1997; Prantzos \& Boissier 2000;
Hernandez \& Cervantes-Sodi 2006), of which our estimate will be an accurate approximation,
for exponential disks having flat rotation curves, as late type galaxies ubiquitously present.

\section{The SDSS sample}

The sample of cluster galaxies used in this work is an updated version of the
PH09 sample (Hwang et al. 2010), using data from the SDSS, Data Release 7
(Abazajian et al. 2009). To
identify the cluster galaxies, we used the Abell cluster catalogue (Abell et al. 1989),
with complementary photometric parameters taken from
Korea Institute for Advanced Study (KIAS) DR7 value-added galaxy catalogues
(Choi et al. 2007; Choi et al. 2010),
and spectroscopic parameters from MPA/JHU and NYU DR7 value-added galaxy catalogues
(Tremonti et al. 2004; Blanton et al. 2005).
The sample contains 730 clusters located within the SDSS survey region
and having known spectroscopic redshifts taken from NED. The position of the cluster
centre is adopted from NED, but replaced with the X-ray determined position, if 
available from the literature.

To establish the membership of galaxies in a cluster, we employed the "shifting gapper" method of Fadda et al. (1996), 
where the radial velocity versus the clustercentric  distance space is used to group galaxies with connection lengths of 
950 km s$^{-1}$ in the direction of the radial velocity and of 0.1 $h^{-1}$Mpc in the direction of the clustercentric 
radius R up to $R=3.5h^{-1}$Mpc. From this procedure, 200 Abell clusters were obtained with at least 10 member galaxies.

Additionally to the galaxies obtained by the "shifting gapper" method, galaxies located
with projected separations of $R_{max} < R < 10r_{200,cl}$ where included to investigate 
the variation of the galactic spin over a wide range of clustercentric radius,
where $R_{max}$ is the largest clustercentric distance of the cluster member galaxies.
An additional constraint was imposed on these galaxies; they have to present a
velocity difference relative to the cluster's systemic velocity of less than 1 000 km s$^{-1}$.

The final sample contains 121 relaxed Abell clusters and 51,842 associated galaxies,
after rejecting the clusters with interacting or merging features, determined from the
galaxy velocity versus clustercentric distance space, and those dynamically young
with the brightest cluster galaxy at large clustercentric distance ($d>0.5 r_{200,cl}$).

The radius $r_{200,{\rm cl}}$, where the mean overdensity drops to 200 times the critical
density of the Universe, is calculated using the formula given by Carlberg et al. (1997):

\begin{equation}
\label{RvirCluster}
r_{200,{\rm cl}}= \frac{3^{1/2}\sigma_{\rm cl}}{10 H(z)},
\end{equation}

where $\sigma _{cl}$ is the velocity dispersion of the cluster, and $H(z)$ the Hubble
parameter given by $H^2(z)=H^2_0 [\Omega_m(1+z)^3 +\Omega_k(1+z)^2+\Omega_\Lambda]$ (Peebles 1993), with
$\Omega_m$, $\Omega_k$, and $\Omega_\Lambda$ the dimensionless density parameters. For this work,
we adopt a flat $\Lambda$CDM cosmology with density parameters $\Omega_{\Lambda}$ = 0.73
and $\Omega_{m}$ = 0.27.

In order to discriminate between elliptical and disc galaxies, we used the prescription of
Park \& Choi (2005) in which early (ellipticals and lenticulars) and late (spirals) types are
segregated in a $u - r$ colour versus $g - i$ colour gradient space and in the concentration index space. 
They tested extensively the selection criteria through direct comparison of visually assigned types
for a large sample of several thousand galaxies. The specific selection criteria can be found in Park \& Choi (2005),
but essentially select as early types, galaxies with red colours, minimal colour gradients and high
concentration indices. 
An additional visual check of the colour images of the galaxies was 
performed to correct misclassifications and assign morphological type
for the galaxies in the DR7 that are not included in KIAS DR7 VAGC.
Throughout the paper we will consider ellipticals and lenticulars as early type 
galaxies, and spirals as late type ones.
Our equation~\ref{Ld} requires the rotational velocity to calculate $\lambda_{d}$,
which is infered from the absolute
magnitude in the r band using the TF relation
$\log V_{d} = -0.135(M_{r} + 21.107) +2.210$ by Pizagno et al. (2007).
To avoid the problem of internal 
absorption in edge-on galaxies (Unterborn \& Ryden 2008; Cho \& Park 2009), and consequently underestimating rotational 
velocities, we limit the sample to spiral galaxies having seeing-corrected
isophotal axis ratios $b/a > 0.6$.

When studying the influence of a nearby galaxy on the value of the spin, the distance between galaxies is
normalized to the virial radius of the nearest neighbour. The nearest neighbour galaxy
of a target galaxy with absolute magnitude M$_{r}$ is the one with the smallest projected
separation distance on the sky to the galaxy, and is brighter than $M_{r} + \Delta M_{r}$
among those in the cluster galaxy sample, with $\Delta M_{r}=0.5$. This particular value is
chosen to assure a large enough sample, so as to get statistically significant results.

We define the virial radius of a galaxy as the projected radius where the mean mass
density within the sphere of radius $r_{vir}$ is 200 times the critical density or
740 times the mean density of the universe;

\begin{equation}
\label{Rvir}
r_{vir}^{3}= \left( \frac{3 }{4 \pi}\right) \left(  \frac{  \gamma L }{200 \rho_{c} }\right)   ,
\end{equation}

with the relative mass-to-light ratios for early type galaxies (ellipticals and lenticulars) twice 
the same ratio for the late types (spirals), $\gamma (early) = 2\gamma (late)$, following Choi et al. (2007), 
who report that the central velocity dispersion of early type galaxies
brighter than $M_{r}=-19.5$, is about $\sqrt{2}$ times that of late types.
Since we adopt $\Omega_{M}=0.27$,
we have $200 \rho_{c}= 200 \overline{\rho}/\Omega_{M}=740\overline{\rho}$, which is almost equal to the virialized
density $\rho_{vir}= 18 \pi ^{2} / \Omega_{m}(H_{0}t_{0})^{2}\overline{\rho}=766\overline{\rho}$, in the
case of a $\Lambda$CDM universe (Gott \& Rees 1975). Finally, we introduced the mean value for the density of
the Universe, $\overline{\rho} = (0.0223 \pm 0.0005)(\gamma L)_{-20}(h^{-1}Mpc)^{-3}$, where $(\gamma L)_{-20}$
is the mass of a late type galaxy with $M_{r}=-20$ (Park et al. 2008).
In this way, the virial radii
of galaxies with $M_{r}=-19.5$, $-$20.0 and $-$20.5 are 260, 300 and 350$h^{-1}$ kpc for early types and
210, 240 and 280$h^{-1}$ kpc for late types, respectively. For more details on the sample, consult PH09 and Hwang et al. (2010).

\section{Results and discussion}

\begin{figure}
\centering
\begin{tabular}{c}
\includegraphics[width=0.475\textwidth]{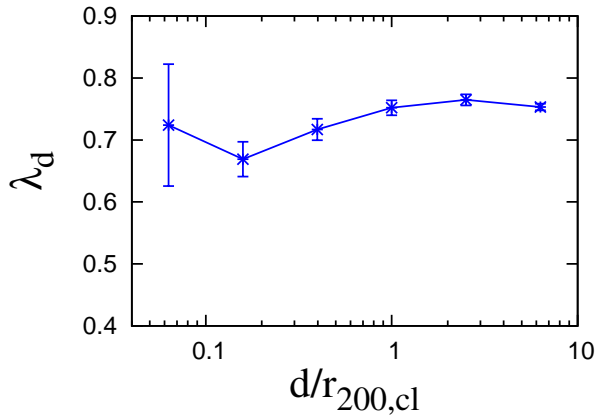} 
 \end{tabular}
\caption[ ]{Median $\lambda_{d}$ value as a function of clustercentric radius for late-type galaxies.}
\label{PlotSpirals}
\end{figure}

In the interest of establishing the dependence of the angular momentum on the 
cluster environment, we first calculate the mean value of $\lambda_{d}$ as a function of the
clustercentric radius normalised to the virial radius of the cluster  $d/r_{200,{\rm cl}}$.
Once having segregated the galaxies into early and late types, we employed
equation~\ref{Ld} to obtain the spin for the disc galaxies in the
sample. Figure~\ref{PlotSpirals} presents the median spin value for disc galaxies
as a function of the clustercentric radius for all clusters in the sample taken together, 
where error bars denote the estimated 1$\sigma$ confidence intervals based on the 
bootstraping re-sampling method. When approaching the center of the cluster,
we note a slight increase of $\lambda_{d}$ and then a systematic drop
for $d \le 2 \times r_{200,cl}$,
with the trend becoming more pronounced on crossing
the virial radius of the clusters, as defined through equation~\ref{RvirCluster}.
To assess the statistical significance of the trend, we performed
a $\chi^{2}$ test taking as null hypothesis that the trend is consistent with
a flat line given by the weighted mean of the data points.
By computing the sum of $\chi^{2}$'s for the data points of
Figure~\ref{PlotSpirals}, weighted by their corresponding associated errors,
we rule out the null hypothesis at a 99.1\% level.

The only possible exception to this trend is in the innermost bin,
with the highest formal uncertainty on the median $\lambda_{d}$ value,  
and where the proximity between galaxies could make the identification
of individual systems difficult. Because, in our sample, the typical
offset of the brightest cluster galaxy from the cluster center is about
$0.1r_{200,cl}$, the uncertainty in determining the cluster center can be
as large as the offset. We will present our results below only for
clustercentric radii larger than $0.1r_{200,cl}$. In spite of a close to 25\% 
uncertainty in each individual estimate of $\lambda_{d}$, the very large
numbers in our sample result in this slight trend being larger than 
our internal uncertainties, it is hence a real detection.

\subsection{Dependence on galaxy mass.}

\begin{figure}
\centering
\begin{tabular}{c}
\includegraphics[width=0.475\textwidth]{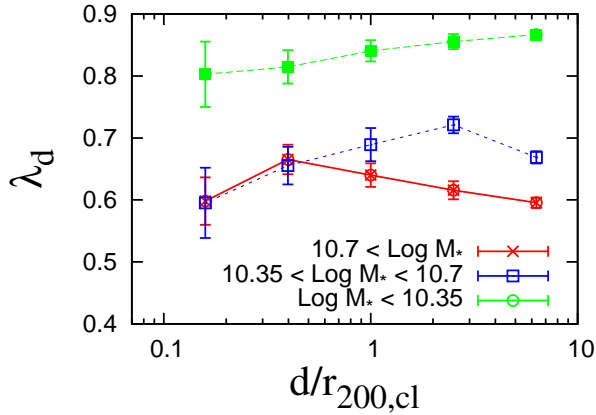}  
\end{tabular}
\caption[ ]{Median $\lambda_{d}$ value as a function of clustercentric
radius for late-type galaxies
in three different stellar mass ($M_{*}$) ranges.}
\label{PlotSpiralsMass}
\end{figure}

It has been reported (Cervantes-Sodi et al. 2008; Berta et al. 2008) that the galactic
spin has a strong dependence on halo mass; typically, massive galaxies conform
a tight population with low $\lambda$ values, while the $\lambda$ distribution of less
massive galaxies presents a higher value with large dispersion. Given
that massive galaxies present low dispersion around their average (low) spin,
we expected that their spin would not show variations with clustercentric
radii, but the case of intermediate and low mass galaxies can be different.
We split the sample of late-type galaxies according to the total stellar galactic mass 
(SDSS-DR7 estimates\footnote{http://www.mpa-garching.mpg.de/SDSS/DR7/Data/stellar mass.html})
into three bins,
to search if the trend shown by the whole sample is followed regardless of
the mass of the galaxies. Figure~\ref{PlotSpiralsMass} presents the spin as a
function of the clustercentric distance for three stellar mass ranges.
Low mass galaxies show monotonically decreasing $\lambda_{d}$ as the clustercentric
radius decreases while intermediate mass galaxies show a peak near $2r_{200,cl}$.
Massive galaxies also show a peak but at much smaller radius of about $0.3r_{200,cl}$.
We perform $\chi^{2}$ tests for these three cases, to evaluate departure
from a constant value. The probabilities that each case is consistent with a constant value
are 82.6\%, 99.2\% and 97.7\%, for the cases of low, intermediate and
massive galaxies respectively. Therefore, the trend seen for the galaxies
in the lowest mass bin is marginal, while those for more massive galaxies
are statistical significant.

Given that the galaxies are immersed in high-density environments, it is likely that
they have undergone high angular momentum removal by tidal or ram stripping,
with a more drastic effect on galaxies residing close to the cluster core,
due to the increasing density towards the center. The effect of these processes
would naturally have a larger influence on low mass galaxies than in massive
ones with deep enough potential wells to retain their material and
preserve their structure.

The possible scenarios describing the evolution of galaxies residing in clusters are
numerous and diverse. Close interactions, collisions and mergers of galaxies are assisted
by the dense environment of galaxy clusters. This kind of processes drastically rearrange
the angular momentum of the galaxies involved (Vitvitska et al. 2002; D'Onghia \& Navarro
2007), although it is still not clear if the net result leads always to an enhance
of the angular momentum of the resultant system (Hiotelis 2008), a systematic
decrease, or if the final result depends on
the particular details of the configuration such as the mass ratio between the galaxies
involved (Hetznecker \& Burkert 2006), their morphology and gas content,
or the orbital parameters of the interaction
(Qu et al., 2010). Specially for the case of central cD galaxies, such dramatic
events are often invoked to explain their structure and formation.
But galaxy-galaxy interactions are not the only processes taking place
in the cluster environment; the interaction with the cluster itself can play a
major role in galaxy evolution. The truncation of dark galactic haloes by the tidal
field of the cluster, a process widely studied (Merritt 1983; Binney \& Tremaine 1988),
can be a source of angular momentum removal, as it stripes material from the outskirts
of the galaxies, leading to a well-defined trend; the closer the galaxies are to the center
of the cluster, the smaller the size of the galaxy (Limousin et al. 2007; Limousin et al.
2009). Particularly, PH09 showed that the size of late-type galaxies starts to decrease
at a clustercentric radius of about $r_{200,cl}$.
But the influence of the cluster tidal field does not affect instantly the structure of the
orbiting galaxies. According to the strength of the cluster force field and the mass
of the galaxy in question, the tidal field will disturb the system pulling the material,
stretching the galaxy and finally stripping unbound material (Miller 1986), producing
the behaviour noticed in Figure~\ref{PlotSpiralsMass}; an initial increase of the
angular momentum given by the gravitational pull, and then the decrease from the subsequent
evolution from non-spherical stretching and compressions, and finally the more external material
being material stripped away.

Repeated interactions with cluster member galaxies can also produce similar effects
on the galaxies orbiting inside the cluster, particularly when they pass through the
central region swarmed with massive galaxies. It is expected that, among the galaxies
gravitationally bound to clusters, the more massive ones have smaller orbital radii.
If the decrease in $\lambda_{d}$ is caused by interactions of galaxies with the cluster
center or with nearby galaxies, the effects will appear at a smaller clustercentric radius
for more massive cluster galaxies. This is actually what we observe in Figure~\ref{PlotSpiralsMass}.
Our analysis in sections 4.2 and 4.3 seems to indicate that the effects of interactions
with the cluster center is more important than the cumulative effects of galaxy-galaxy
interactions in changing $\lambda_{d}$, at least in the case of massive clusters.

Gunn \& Gott (1972) proposed that the ram pressure of the intracluster medium might
sweep away gas from the galaxies if the force of the ram-pressure stripping exceeds the 
restoring gravitational force of the galaxy, implying the presence of a stripping radius.
In this sense, small galaxies will not be able to retain their gas beyond this
stripping radius, specially at the center of the clusters where the ram pressure is
larger, losing material with high angular momentum, while galaxies large enough
will not present this truncation. This mechanism of material removal explains satisfactorily
many features observed in cluster galaxies, such as the morphology-density relation
(van der Wel et al. 2010), the star formation (Kronberger, Kapferer, Ferrari,
Unterguggenberger \& Schindler 2008; Weinmann et al. 2009), or the Butcher-Oemler effect
(Abadi, Moore \& Bower 1999). This mechanism of material removal will specially affect
galaxies rich in gas, truncating the gas disc, and in consequence no stars will be formed
beyond this radius.

Although ram-pressure stripping may account for many features of cluster
galaxies, tidal interaction between galaxies and substructure within the cluster
are often invoked to explain the same observed features, being able to significantly
change the structure and activity of galaxies (Moss \& Whittle 2000; Gnedin 2003).

Whether the galaxy-galaxy interactions or the cluster-galaxy interaction dominate
in the picture of galaxy transformations, is still not clear. Given that the local galaxy
density and the cluster mass density are correlated for cluster galaxies (Dom\'inguez,
Muriel \& Lambas 2001), it is difficult to distinguish the effect of local and
large scale density on observed galaxies, besides, the frequency of galaxy-galaxy interactions
depends on the clustercentric radius, increasing the uncertainty on which interactions
are responsible for the galaxy transformations.

As mentioned before in the introduction, Park, Gott \& Choi (2008)
and Park \& Choi (2009) have found that galaxy
properties like morphology and luminosity depend on the distance to the nearest
neighbour galaxy, even at fixed large-scale density, specially when the galaxy is
located within the virial radius of its nearest neighbour, with a major role played
by the morphology of the neighbour. Regarding the spin parameter,
in Cervantes-Sodi, Hernandez \& Park (2010), we showed that the angular momentum is
also sensitive to galaxy-galaxy interactions,
leading to a gradual decrease in the values
of $\lambda$, as soon as the galaxies cross into their virial radii. This being so,
the trends exhibited by our present sample could arise simply from ongoing
galaxy-galaxy interactions, and not directly due to the interaction with the
cluster itself, or could be a joint effect form the local density plus the neighbour
environment, as proposed by PH09. PH09 found that the structural and kinematic
parameters of late-type galaxies such as the concentration index and the central
velocity dispersion depend mainly on the clustercentric radius, but the star
formation activity indicators like the u-r colour and the equivalent width of
H$\alpha$ line, depend predominantly on the neighbour environment, where the neighbour
environment means both neighbour morphology and the mass density attributed to the
neighbour. They concluded that transformations of late-type
galaxies in the clusters are not due to a single mechanism, but due to both;
hydrodynamic and gravitational processes.

In the following section we will investigate if the trends found in this section
are due to the proximity of a nearby companion or if it is a compound effect of
long term galaxy-galaxy interactions plus the influence of the cluster as a whole.

\subsection{Interacting and non-interacting galaxies}

\begin{figure}
\centering
\begin{tabular}{c}
\includegraphics[width=.475\textwidth]{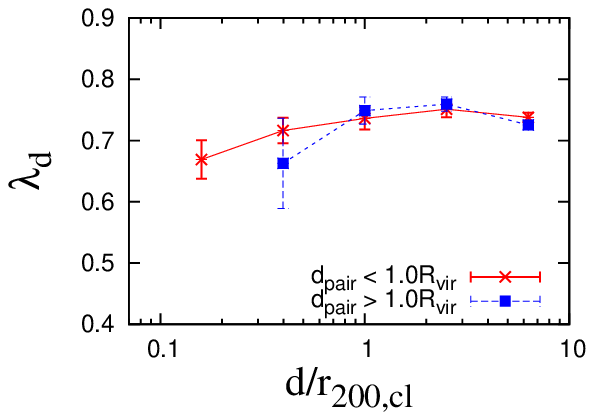} \\
\includegraphics[width=.475\textwidth]{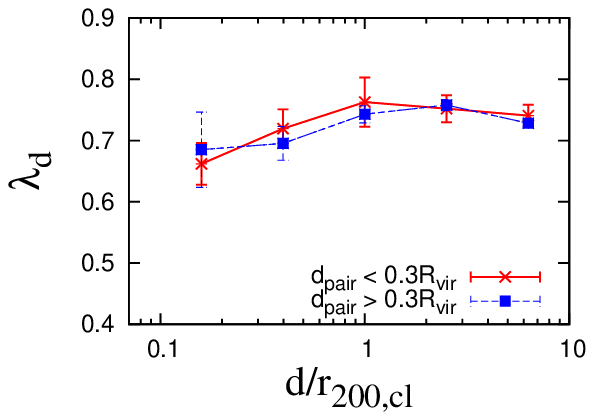}
\end{tabular}
\caption[ ]{Median $\lambda_{d}$ value as a function of clustercentric radius for late-type galaxies,
for galaxies segregated according to the distance of the nearest neighbour galaxy, using two
threshold values;
 \textit{top panel:} 1.0R$_{vir}$,
\textit{bottom panel:} 0.3R$_{vir}$.
}
\label{RvirPlot}
\end{figure}

We have shown how the median value of $\lambda_{d}$ tends to decrease as we approach the center
of the cluster. Simultaneously, both the cluster density and the galaxy-galaxy
interactions, that tend to reduce the spin of the involved systems, increase.

To clarify if the behaviour found is coming from the increase of galaxy-galaxy interactions
toward the center of the clusters, we study separately those galaxies in interaction, i. e.
with a separation distance within one virial radius of the neighbour galaxy, $1.0R_{vir}$,
and those relatively isolated, with a separation distance to the nearest neighbour,
$d>1.0R_{vir}$. The comparison between the two groups is presented in Figure~\ref{RvirPlot},
top panel. If the spin dependence on the clustercentric
radius comes form ongoing galaxy encounters in the inner regions, we would
expect that galaxies currently in interaction would present the reported trend,
while those with no close companion should not.
This is not the case; both classes, the interacting and
non-interacting galaxies, follow the same trend irrespectively of the presence of a
nearby companion. In this case we performed $\chi^{2}$ tests with the null hypothesis
being that the two sub-samples follow the same trend, taken as the weighted mean trend
of the two sub-samples. The probability under the null hypothesis that the two samples
follow the same trend is 71.1\% and 80.2\% for galaxies with the nearest neighbour galaxy
within $d>1.0R_{vir}$ and $d<1.0R_{vir}$ respectively. We have two different values
for the $\chi^{2}$ tests because the average trend is obtained from the weighted mean value,
for each bin, of the two sub-samples, and each sub-sample has different number of galaxies
in each bin. As it is apparent by eye, both sets show the same trend.

The number of galaxies with
$d_{pair}>1.0R_{vir}$ drops drastically for the inner regions, as the number density of galaxies
increases, which makes it impossible to inspect the most inner parts of the clusters. 

\begin{figure}
\centering
\begin{tabular}{c}
\includegraphics[width=0.475\textwidth]{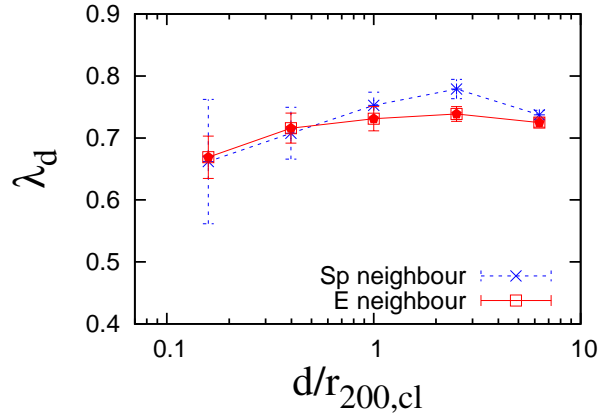}  
\end{tabular}
\caption[ ]{Median $\lambda_{d}$ value as a function of clustercentric radius for late-type galaxies,
segregated according to the nearest neighbour morphology.}
\label{MorphPlot}
\end{figure}

In Cervantes-Sodi, Hernandez \& Park (2010), we showed that the perturbation on the
spin parameter produced by the presence of a nearby companion is more tangible for
$d_{pair}<0.5R_{vir}$. Choosing a value smaller than $0.5R_{vir}$, will allow
us to explore inner regions with galaxies still not showing clear features of
galaxy-galaxy interactions. We choose $0.3R_{vir}$ to segregate galaxies with
clear interaction features, and present the result in Figure~\ref{RvirPlot}, bottom panel.
With this new cut, we can explore the region close to the cluster centre with
"non-interacting" galaxies, and we clearly see that the trend persist regardless of
the small scale environment. The $\chi^{2}$ test for the two samples to follow one common
trend derived as the mean of both, accepts the null hypothesis for a common trend
with a level of significance $>$ 91.1\% for both sub-samples. This result implies that even
when galaxy-galaxy interactions play an important role on modifying the galactic
spin, the larger scale environment of the cluster plays also an important systematic role
in changing the structure of the systems, as early suggested by PH09.

As a final test on the joint role played by galaxy-galaxy interactions and galaxy-cluster
interactions, we contrast the spin-clustercentric radius relation according to the morphology
of the nearest neighbour. In Cervantes-Sodi et al. (2010),
we noticed that the response to galaxy-galaxy
interactions depends on the morphology of the nearest neighbour, the case of an
interaction with a late-type galaxy showing much more significant disturbances in the
spin value, probably due to the strong hydrodynamical effects of the interactions of
two gaseous discs, specially in very close encounters, with $d_{pair}<0.1R_{vir}$.
We plot the median $\lambda_{d}$ value as a function of $d/r_{200,cl}$ for galaxies
segregated according to the morphological type of their nearest neighbour
(Figure~\ref{MorphPlot}). The trend for both cases is similar showing
a gradual decrease of $\lambda_{d}$ as the galaxies approach the cluster centre and cross
the virial radius. A $\chi^{2}$ test comparing the reported trends with an average weighted
one derived from the two originals, assigns a 60.9\% confidence level
for the null hypothesis of both samples following the same trend.
Examining the galaxies entailed in obtaining the median
value of those bins, we notice that all of them are located within 0.5$R_{vir}$
of the nearest neighbour, reinforcing our previous conclusion, suggesting galaxy-cluster 
interactions as driving the observed trend.

\subsection{Dependence on cluster mass.}

\begin{figure}
\centering
\begin{tabular}{c}
\includegraphics[width=0.475\textwidth]{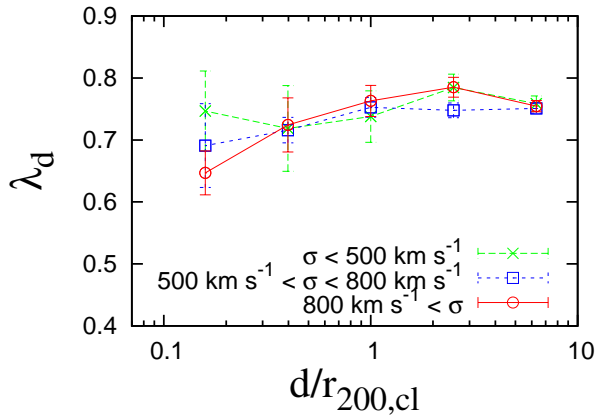}  
\end{tabular}
\caption[ ]{Median $\lambda_{d}$ value as a function of clustercentric radius for late-type galaxies,
segregated according to the velocity dispersion of the cluster.}
\label{ClusterPlot}
\end{figure}

One particularly interesting question we can ask is whether or not this trend we have found
depends on the mass of the cluster. If this trend is imposed through the interactions with
the cluster environment, the bigger the galaxy cluster, the stronger the interaction with
its potential will be. In this sense, we expect a clearer trend when dealing with galaxies
immersed in massive galaxy clusters. The result of dividing our sample into three cases,
according to the velocity dispersion of the hosting cluster, is shown in Figure~\ref{ClusterPlot},
where the decrease of $\lambda_{d}$ for small clustercentric radius is present only
for the case of galaxy clusters with $\sigma >$ 500 km s$^{-1}$, indicating a dependence on the
galaxy cluster mass, and a possible clue of the galaxy-cluster interaction as the process
responsible for establishing the angular momentum-clustercentric radius relation,
although we must be cautious given the large error bars involved.
Surprisingly, this cut in $\sigma$ coincides with the value reported by Poggianti et al.
(2006); who found that the star formation activity is drastically suppressed for galaxies
residing in massive enough clusters with $\sigma >$ 550 km s$^{-1}$.

Following the same methodology as in Section 4.1,
we performed $\chi^{2}$
tests to check the significance of the trends, with the null hypothesis being a lack
of dependence of $\lambda_{d}$ on the clustercentric radius. We can reject the null hypothesis
at a 99.1\% level for the case of the  high velocity dispersion clusters,
while the trend for galaxies in low mass clusters is consistent with no radial dependence at a 74.2\%
level of significance. Finally, for the intermediate mass clusters, the null hypothesis
of there being no trend is consistent with the data.

\begin{figure}
\centering
\begin{tabular}{c}
\includegraphics[width=.475\textwidth]{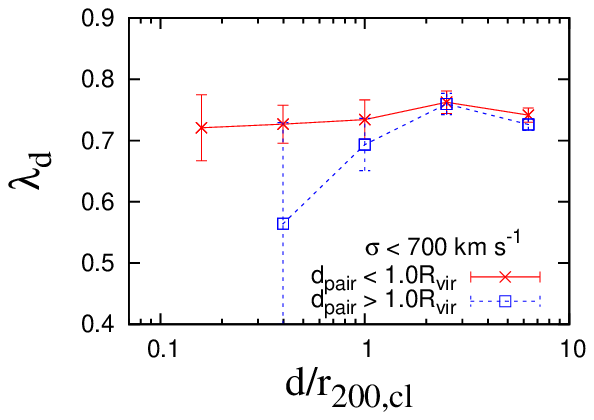} \\
\includegraphics[width=.475\textwidth]{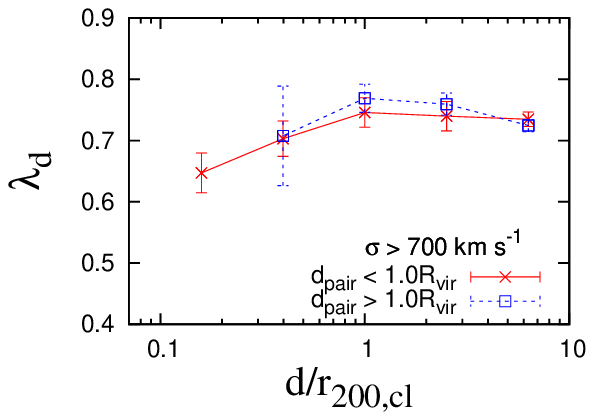}
\end{tabular}
\caption[ ]{Median $\lambda_{d}$ value as a function of clustercentric radius for late-type galaxies,
for galaxies segregated according to the velocity dispersion of the hosting cluster;
low mass galaxy clusters with $\sigma <$ 700 km s$^{-1}$ (top panel), and
massive clusters with $\sigma >$ 700 km s$^{-1}$ (bottom panel).
}
\label{Cluster-Neigh}
\end{figure}

In the preceding subsection, we showed the joint effect that galaxy-galaxy and
galaxy-cluster interactions have in modifying the spin of disc galaxies. Given that
galaxies immersed in massive galaxy clusters present the strongest
$\lambda_{d}$ dependence on the clustercentric radius,
we are interested in pondering the importance of each of these
processes for different cases. Figure~\ref{Cluster-Neigh} presents two different cases,
in the upper panel galaxies immersed in low mass clusters ($\sigma <$ 700 km s$^{-1}$),
the lower panel with galaxies of massive clusters ($\sigma >$ 700 km s$^{-1}$), all of
them segregated according to the distance to the nearest neighbour. For the case of massive
clusters, the trend with the clustercentric radius is clear for all the galaxies, regardless
of the presence of a nearby companion, indicating an origin for the trend independent
of recent galaxy interactions, rather grounded on galaxy-cluster interactions. We confirm
this conclusion by a $\chi^{2}$ test for the two sub-samples of Figure~\ref{Cluster-Neigh}
lower panel, following the same trend, drawn as the average weighted
one derived from the two originals, with a level of significance $>$ 86.9\%.
The case of low mass clusters is completely different, where
interacting galaxies depart from the trend of non-interacting galaxies as can be
seen in Figure~\ref{Cluster-Neigh} upper panel, with a
significance level $<$ 67.2\% for the two classes to follow the same trend, derived as
the average weighted one of the two originals. 
Clearly, in these systems the gravitational interaction with the cluster
potential is diminished due to its low mass, and the effect of ongoing interactions with
close neighbours dominates.

For interacting galaxies in low mass clusters, distance to the cluster centre
is not a factor in fixing the internal dynamics or 'type', as galaxy-galaxy
interactions are driving morphological changes, being the overall cluster
effects secondary. On the other hand, for massive clusters the trend for
diminishing $\lambda_{d}$ as cluster-centric distances drop, holds identically
both for interacting and non-interacting galaxies, being galaxy-cluster
interactions responsible for the leading dynamical evolution of galaxies.

\section{Conclusions}

Using a large sample of cluster galaxies, we study the dependence of the galactic
angular momentum on the environment, accounted for by the clustercentric radius, and found
a slight increase of the median value of our dimensionless angular momentum parameter,
as we approach the clusters and a systematic decrease as we approach the central region.
This $\lambda_{d}$-clustercentric radius
relation is not equally followed by all galaxies; only low and intermediate mass
galaxies present this dependence on the clustercentric radius, while the spin
of massive galaxies is more robust. A possible interpretation
for the different response of massive galaxies could be the relative
stability of these systems against perturbations, such as galaxy-galaxy or
galaxy-cluster interactions, harassment or ram pressure stripping,
given the depth of their gravitational potentials, and their different formation
time; as recently pointed out by Huertas-Company et al. 2009, it seems that these
galaxies experience little evolution in the cluster environment, as they were
probably formed at higher redshifts before entering the cluster in infalling
groups.

Given the increased proximity of galaxies in the central parts of galaxy
clusters, the trend found could arise directly from galaxy-galaxy
interactions in this environment.
To test if the trend is established only by ongoing interactions with
neighbouring galaxies, we tested if the galaxies followed the same trend
regardless of the presence of nearby companions. To this end, we split the
sample into galaxies having their nearest neighbour galaxy within one
pre-establish distance, given in terms of the virial radius of the
neighbour (1.0$R_{vir}$ or 0.3$R_{vir}$). The result is very similar
for both cases, the trend is shown by interacting and
non-interacting galaxies, and regardless of the morphology of the nearest
neighbour, with minor differences within the statistical
errors of reported values.

The trend seems to be followed only in massive enough galaxy clusters, where
gravitational interactions of the galaxies with the cluster potential prevail
over galaxy-galaxy interactions, which is not the case for low mass clusters. 
Below $\sigma=700$ km/s for the clusters, one is dealing with an effective large group,
in regards to galactic morphology transformations, which are mostly determined
by galaxy-galaxy interactions, and only the case of non-interacting galaxies
present the trend with the clustercentric radius.

The primary point we want to emphasize is the clear dependence of our $\lambda_{d}$
estimate of the galactic spin on the clustercentric radius, even at fixed stellar mass
of the galaxies, with typically low $\lambda_{d}$
values in the inner regions of galaxy clusters. The trend is dependent on
the mass of the galaxies, shown only in low and intermediate mass galaxies,
while the $\lambda_{d}$ value of massive galaxies show essentially low values when compared
with less massive galaxies. The influence of ongoing
galaxy-galaxy interactions on the spin is present, but can not account
for the trend with the clustercentric radius in massive galaxy clusters
with $\sigma>700$ km/s, where galaxy-cluster interactions dominate and drive 
morphological transformations.

Having an objective and quantitative parameter to assess galactic type
and its transformations, allows us to complement and better understand qualitative
studies based on less objective type assessments, such as measures of E/S fractions.
The large and high quality sample of clusters and galaxies used permits to
meaningfully dissect the trends observed and hence identify the causes, sorting out
galaxy-galaxy interactions from galaxy-cluster ones.

\section*{Acknowledgments}

The authors acknowledge the thorough reading of the original manuscript by the 
anonymous referee, as helpful in reaching a clearer and more complete final version.

C.B.P. was supported by the National Research Foundation of Korea (NRF)
grant funded by the Korea government (MEST) (No. 2009-0062868).

Funding for the SDSS and SDSS-II has been provided by the Alfred P. Sloan
Foundation, the Participating Institutions, the National Science
Foundation, the U.S. Department of Energy, the National Aeronautics and
Space Administration, the Japanese Monbukagakusho, the Max Planck
Society, and the Higher Education Funding Council for England.
The SDSS Web Site is http://www.sdss.org/.

The SDSS is managed by the Astrophysical Research Consortium for the
Participating Institutions. The Participating Institutions are the
American Museum of Natural History, Astrophysical Institute Potsdam,
University of Basel, Cambridge University, Case Western Reserve University,
University of Chicago, Drexel University, Fermilab, the Institute for
Advanced Study, the Japan Participation Group, Johns Hopkins University,
the Joint Institute for Nuclear Astrophysics, the Kavli Institute for
Particle Astrophysics and Cosmology, the Korean Scientist Group, the
Chinese Academy of Sciences (LAMOST), Los Alamos National Laboratory,
the Max-Planck-Institute for Astronomy (MPIA), the Max-Planck-Institute
for Astrophysics (MPA), New Mexico State University, Ohio State University,
University of Pittsburgh, University of Portsmouth, Princeton University,
the United States Naval Observatory, and the University of Washington.

\end{document}